\documentclass[conference]{IEEEtran}
\IEEEoverridecommandlockouts
\usepackage{cite}
\usepackage{amsmath,amssymb,amsfonts}
\usepackage{algorithmic}
\usepackage{graphicx}
\usepackage{textcomp}
\usepackage{xcolor}
\def\BibTeX{{\rm B\kern-.05em{\sc i\kern-.025em b}\kern-.08em
    T\kern-.1667em\lower.7ex\hbox{E}\kern-.125emX}}
\begin{document}

\IEEEoverridecommandlockouts
\IEEEpubid{\makebox[\columnwidth]{978-1-5386-5541-2/18/\$31.00~\copyright2021 IEEE \hfill} \hspace{\columnsep}\makebox[\columnwidth]{ }}
\title{Bunching Dynamics of Buses in a Loop}
\IEEEpubidadjcol

\author{
\IEEEauthorblockN{Luca Vismara}
\IEEEauthorblockA{\textit{Interdisciplinary Graduate Programme} \\
\textit{Nanyang Technological University} \\
Singapore, Singapore \\
Email: vism0001@e.ntu.edu.sg \\
ORCID: 0000-0002-5216-7975
}
\and
\IEEEauthorblockN{Vee-Liem Saw}
\IEEEauthorblockA{
\textit{School of Physical and Mathematical Sciences} \\
\textit{Nanyang Technological University}\\
Singapore, Singapore \\
Email: vee-liem@ntu.edu.sg\\
ORCID: 0000-0003-3621-3799}
\and
\IEEEauthorblockN{Lock Yue Chew}
\IEEEauthorblockA{
\textit{School of Physical and Mathematical Sciences} \\
\textit{Nanyang Technological University}\\
Singapore, Singapore \\
Email: lockyue@ntu.edu.sg\\
ORCID: 0000-0003-1366-8205}
}

\maketitle

\begin{abstract}
Bus bunching is a curse of transportation systems such as buses in a loop. Here we present an analytical method to find the number of revolutions before two buses bunch in an idealised system, as a function of the initial distance and the crowdedness of the bus stops. We can also characterise the average waiting time for passengers as the buses bunch. The results give a better understanding of the phenomenon of bus bunching and design recommendations for bus loops.
\end{abstract}

\begin{IEEEkeywords}
transportation, buses, dynamical systems, bus bunching, waiting time
\end{IEEEkeywords}

\section{Introduction}
Bunching is a problem that plagues transportation systems from trains to buses \cite{saw2019}. Bus bunching happens when two or more consecutive buses arrive at a bus stop at the same time, moving as a platoon. Without an active control \cite{Chew2019} \cite{Daganzo2009} \cite{saw2019intelligent}  \cite{Abkowitz1984} \cite{rossetti1998} \cite{Hickman2001} \cite{fu2002} \cite{cats2011} \cite{bartholdi2012} \cite{Moreira-Matias2016} \cite{quek2020analysis} \cite{saw2019} \cite{Wang2020} or a specific design of bus stops \cite{vismara21} \cite{Fu2003} \cite{LEIVA20101186} \cite{Yavuz2010} \cite{Chiraphadhanakul2013} \cite{chen2015design} \cite{larrain2015generation} \cite{soto2017new}, bunching is inevitable \cite{saw2019}  \cite{Chew2019} \cite{Newell1974} and it causes longer waiting time and delays.

Hereby we present a formalism to study bus bunching in an idealised loop with different settings, to individuate the important variables that affect the phenomenon and the timescale at which it happens. 
In bus loops, the delay or advance of buses at a given point in time is carried over to the next loop.  Bus loops are therefore more susceptible to bunching compared to bus lines where buses are removed from the system when they reach the end of the line and reintroduced according to a given schedule at the beginning of the bus.

In section \ref{sec:loop} we present three cases for a bus loop: the simplest case of two buses and a single bus stop (\ref{ssec:onestop}), the extension to multiple origin bus stops (\ref{ssec:Mstops}) and the case for one origin and one destination bus stop, explicitly accounting for alighting (\ref{ssec:alight}). From those results, we show that introducing more bus stops (at constant total demand of passengers) delay bunching. In section \ref{sec:waiting}, we link the main quantity of interest for bus bunching, the distance between buses, with the waiting time for passengers at bus stops showing that indeed staggered buses minimise the waiting time.
The final part is a summary of the results and limitations of our approach.

\section{Bunching in a bus loop} \label{sec:loop}
Let us consider a bus loop with $M$ arbitrarily positioned bus stops and 2 buses. Each bus can board or alight $l$ passengers per unit time. Passengers arrive at each bus stop at the rate of $s$ per unit time. It is convenient to describe this system in terms of the ratio between those quantities: $k = s/l$ with $0 \le k < 1$. The second inequality ensures that a bus is able to serve a bus stop by boarding more passengers per unit time than the number of new passengers per unit time arriving while the bus is boarding. 
In empirical systems, the value of $k$ is typically very small, $k < 0.1$ \cite{saw2019no}.
If $k \ge 1$ a bus would not move from a bus stop because it will never finish boarding. We have here assumed that the bus has infinite carrying capacity.
We define $T$ as the time taken by a bus to complete the loop without stopping at bus stops.
To study bunching, we consider the quantity $\Delta_n$ as the shortest distance between the two buses at the beginning of the $n$th loop. In this paper, a loop begins when the first bus reaches the first bus stop. The initial distance is $\Delta_0 \le T/2$.

\subsection{One bus stop}\label{ssec:onestop}
\begin{figure}[htbp]
\centerline{\includegraphics[width=0.45\textwidth]{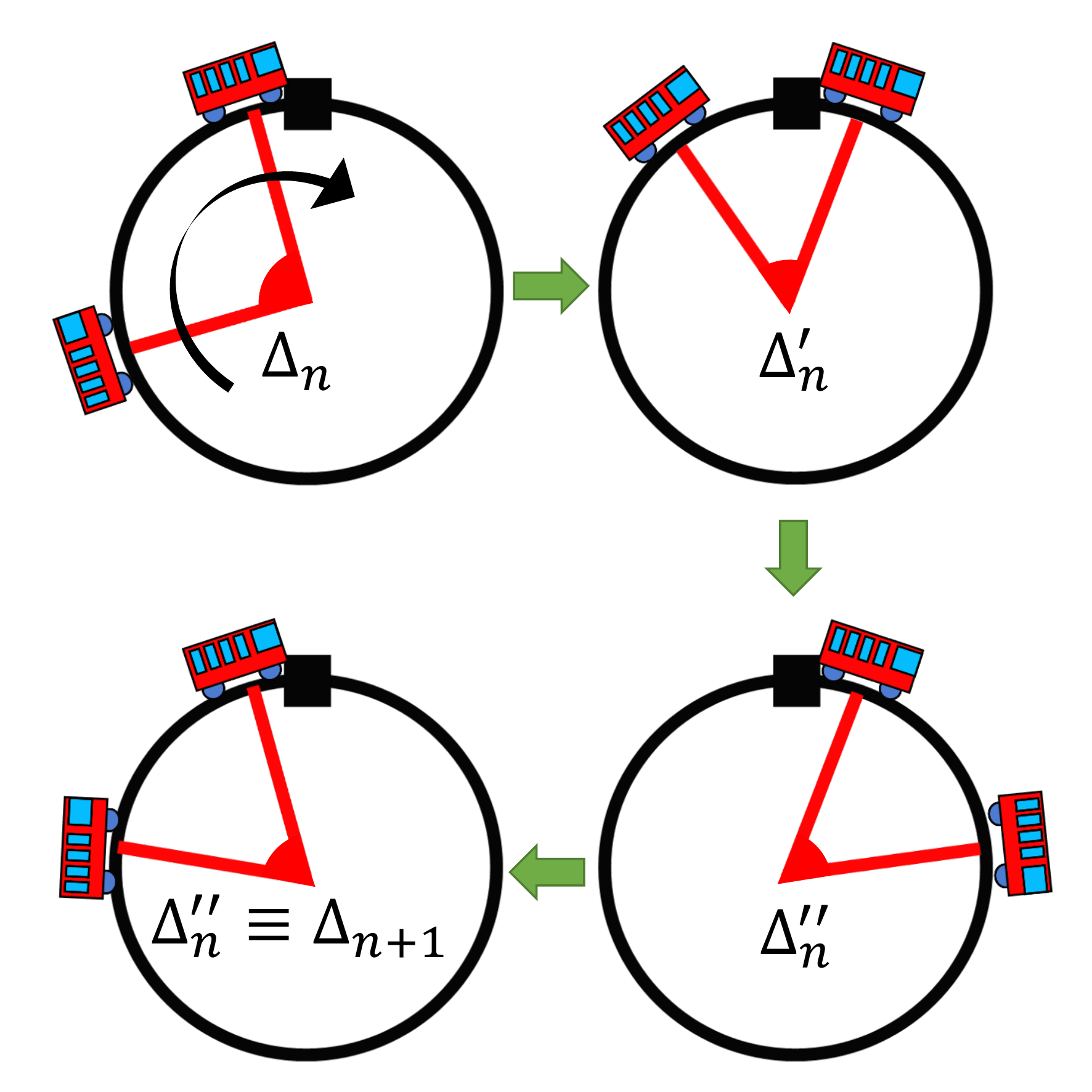}}
\caption{Dynamics of the minimum distance between buses $\Delta_n$ during a loop in the case of two buses and one origin bus stop. The mathematical expression of $\Delta_n$ and $\Delta_{n+1}$ are described by Eqs. \eqref{eq:delta_n_1} and \eqref{eq:dwell_n_1}.}
\label{fig}
\end{figure}

The simplest case to study is one bus stop $M=1$ served by two buses where passengers only board. Following the scheme in figure \ref{fig}, it is possible to find how the distance between the two buses changes between the $n$th loop ($\Delta_n$) and the subsequent loop ($\Delta_{n+1}$).
\begin{equation}\label{eq:delta_n_1}
    \left\{
    \begin{aligned}
        \Delta_n' &= \Delta_n - \tau_n^{(1)} \\
        \Delta_n'' &= \Delta_n' + \tau_n^{(2)} = \Delta_{n+1}.
    \end{aligned}
    \right.
\end{equation}
We denote the dwell time at the bus stop for the $i$th bus at the $n$th loop with $\tau_n^{(i)}$, with the convention that the index 1 is for the bus ahead and 2 is for the bus behind. 
For the distances, we use $\Delta_n'$ and $\Delta_n''$ to indicate the distance between bus 1 and bus 2 within the same loop after the first (prime) and the second (double prime) bus serve the bus stop, as in figure \ref{fig}. The expression for the dwell time is
\begin{equation}\label{eq:dwell_n_1}
    \left\{
    \begin{aligned}
        \tau_n^{(1)} &= \frac{k}{1-k} \left( \bar{T}_n - \Delta_n \right) \\
        \tau_n^{(2)} &= \frac{k}{1-k} \Delta_n'\\
        \bar{T}_n &= T.
    \end{aligned}
    \right.
\end{equation}

Dwell time is proportional to the number of passengers waiting since boarding happens at a rate of $l$ passengers per unit time.
In our model, passengers arrive at a constant rate $s$ from the time the previous bus leaves the bus stop until the current bus boards the last passengers. For $\tau_n^{(2)}$, $s \Delta_n'$ passengers are waiting at the bus stop when the bus arrives, but $s \Delta_n' \times s/l + s \Delta_n' \times s^2/l^2  + \cdots s \Delta_n' \times s^j/l^j + \cdots$ new passengers arrive while boarding. The expressions in Eqs. \eqref{eq:dwell_n_1} follow from the definition of $k=s/l$ and the geometric series. The quantity $\bar{T}_n$ refers to the difference in time between bus 2 leaving the bus stop and bus 2 reaching the same bus stop again. In this case, it is equal to $T$ because there are no other stops in between. Similarly to the calculation for $\tau_n^{(2)}$, the number of passengers waiting when bus 1 arrives at the bus stop is $s (T - \Delta_n)$ and by considering passengers arriving while the bus is boarding, the total dwell time is $\tau_n^{(1)} = k/(1-k) (T - \Delta_n)$.

By substituting Eqs. \eqref{eq:dwell_n_1} in Eqs. \eqref{eq:delta_n_1} the distance between the two buses at the $n$th loop is:
\begin{equation}\label{eq:delta_recurrent_1}
    \Delta_{n+1} = \Delta_n - \tau_n^{(1)} + \tau_n^{(2)} = \frac{\Delta_n}{\left(1-k\right)^2} - \frac{k}{\left(1-k\right)^2} T.
\end{equation}
The recurrence equation can be solved with the initial value $\Delta_0$:
\begin{equation} \label{eq:delta_n_sol}
    \Delta_n = \frac{T}{2-k} - \frac{T - \Delta_0 \left(2-k\right)}{2-k} \left(\frac{1}{\left(1-k\right)^2}\right)^{n}.
\end{equation}
Bunching occurs when the distance between the buses is zero. From Eq. \eqref{eq:delta_n_sol}, the number of loops $n^*$ for the buses to bunch starting from a distance $\Delta_0$ is the solution of $\Delta_{n^*} = 0$:
\begin{equation}\label{eq:n_1}
    n^* = \frac{\log{\left(1 - \frac{\Delta_0}{T} \left(2-k\right)\right)}}{\log{\left(\left(1-k\right)^2\right)}}.
\end{equation}
With the constraint of $\Delta_0 \le T/2$, as $\Delta_n$ is defined as the shortest distance between buses, the number of loops before bunching is maximised by $\Delta_0 = T/2$. In this case, $n^* = (\log (k/2))/(2 \log (1-k))$ so the number of loops  before bunching is a monotonically decreasing for $0 \le k < 1$.

\subsection{\textbf{$M$}bus stops}\label{ssec:Mstops}
In this subsection, we present the result for $M$ arbitrarily positioned bus stops where only boarding is considered. The idea is the same as in the previous case: Eq. \eqref{eq:delta_recurrent_1} will contain two terms $\tau_n$ for each of the $M$ bus stops:
\begin{equation}\label{eq:delta_recurrent_M}
    \Delta_{n+1} = \Delta_n - \sum_{j=1}^M \left( \tau_n^{(1, j)} - \tau_n^{(2, j)} \right),
\end{equation}
where the terms in the sum $\tau_n^{(i, j)}$ are the dwell times of bus $i$ at bus stop $j$ at the $n$th loop. For bus stops with the same $k$, the expression in Eq. \eqref{eq:delta_recurrent_M} is equivalent to iterating Eq. \eqref{eq:delta_recurrent_1} for $M$ times. The number $n^*$ of loops before bunching, in the case of $M$ bus stops with identical $k$, is
\begin{equation}\label{eq:bunching_n_M}
    n^* = \frac{1}{M} \frac{\log{\left(1 - \frac{\Delta_0}{T} \left(2-k\right)\right)}}{\log{\left(\left(1-k\right)^2\right)}}.
\end{equation}
An interesting case of practical concern in designing a bus system is deciding whether to split a crowded bus stop into less crowded ones. With a fixed combined demand of $S$ passengers per unit time arriving at all of the bus stops, would subdividing the demand between multiple bus stops delay bunching? Considering two perfectly staggered buses $\Delta_0=T/2$ and $M$ identical bus stops, each with $s=S/M$ passengers per unit time arriving, hence each with $k=(S/M)/l = K/M$ in Eq. \eqref{eq:bunching_n_M}, bunching occurs after
\begin{equation}\label{eq:n_M}
    n^* = \frac{1}{M} \frac{\log{\left(\frac{K}{2 M} \right)}}{\log{\left(\left(1-\frac{K}{M}\right)^2\right)}}.
\end{equation}
From the equation above it can be shown that $\partial n^* / \partial M > 0$ for any valid $K$ and for $M>1$, so adding more bus stops (while keeping the total number of passengers arriving in the bus loop constant) delays bunching. In the limit of an infinite number of bus stops, $\lim_{M \to \infty} n^* = \infty$. Expanding the expression in Eq. \eqref{eq:n_M} in the limit of big $M$ shows a logarithmic growth of the number of loops before bunching: $n^*_{\text{approx}} = \log{((2M)/K)}/(2 K)$, with $\lim_{M \to \infty} (n^* - n^*_{\text{approx}}) \to 0$.

\subsection{Boarding and alighting}\label{ssec:alight}
In section \ref{ssec:onestop} we have not considered alighting for simplicity. It is possible to write an approximate expression (an upper limit) for the number of loops before bunching when alighting is explicitly considered. 

Let us consider the configuration of two bus stops in a loop: an origin bus stop, where passengers arrive at a rate of $s$ per unit time and a destination bus stop where passengers alight. Boarding and alighting happen at different bus stops. The framework can model general bus stops (where boarding and alighting occur sequentially) by placing a destination bus stop just before an origin bus stop: passengers alight first and then board \cite{saw2020chaos} \cite{vismara21}. We do not expand on this extension for $M>2$ bus stops in this work because the approximation gives less accurate results in this regime, hence we will study the case for one origin and one destination bus stops.
Following the same idea as in Eq. \eqref{eq:delta_n_1}:
\begin{equation}\label{eq:delta_n_M}
    \left\{
    \begin{aligned}
        \Delta_n' &= \Delta_n - \tau_n^{(1)} \\
        \Delta_n'' &= \Delta_n' + \tau_n^{(2)} \\
        \Delta_n''' &= \Delta_n'' - \bar{\tau}_n^{(1)} \\
        \Delta_{n+1} &= \Delta_n' + \bar{\tau}_n^{(2)}.
    \end{aligned}
    \right.
\end{equation}
We denote the boarding time at the origin bus stop for the $i$th bus at the $n$th loop with $\tau_n^{(i)}$. The time spent alighting at the destination bus stop is described by $\bar{\tau}_n^{(i)}$. The ``prime'' notation is used to differentiate between changes of distance between the buses within the same loop $n$ as in the previous case. See Fig. \ref{fig} for a visual explanation.
We make the reasonable assumption that boarding and alighting happen at the same rate of $l$ passengers per unit time. Under this premise, $\tau_n^{(i)} = \bar{\tau}_n^{(i)}$ and the system is characterised by the dimensionless rate $k=s/l$ used in the previous sections.
Similarly to Eq. \eqref{eq:dwell_n_1} we can express the dwell times as a function of $k$ and the distance of buses:
\begin{equation}\label{eq:dwell_n_M}
    \left\{
    \begin{aligned}
        \tau_n^{(1)} &= \frac{k}{1-k} \left( \bar{T}_n - \Delta_n \right) \\
        \tau_n^{(2)} &= \frac{k}{1-k} \Delta_n'\\
        \bar{\tau}_n^{(1)} &= \tau_n^{(1)}\\
        \bar{\tau}_n^{(2)} &= \tau_n^{(2)}\\
        \bar{T}_n &= T + \bar{\tau}_{n-1}^{(2)}  \approx T.
    \end{aligned}
    \right.
\end{equation}
The difference with the previous case is that $\bar{T}_n$ is only approximately equal to $T$. 
The quantity $\bar{T}_n$ is the difference in time between bus 2 leaving the origin bus stop (at the $(n-1)$th loop) and bus 2 reaching the same bus stop again at the $n$th loop. 
The variable $\bar{T}_n$ depends on the dwell time of bus 2 at the destination bus stop at the $(n-1)$th loop $\bar{\tau}_{n-1}^{(2)}$, which in turn is the dwell time of bus 2 at the origin bus stop, which is a function of $\bar{T}_{n-1}$, hence the exact expression does not have a closed form given $\Delta_0$. 
The simplification $\bar{T}_n \approx T$ underestimates the real value, so the number of loops before bunching $n^*$ obtained in this setting is an upper limit of the real value. The approximation is valid when $\bar{\tau}_{n-1}^{(2)}$ is small, hence either small $k$ (low demand) or small $\Delta_n$ (the buses are close to bunching).
By combining Eqs. \eqref{eq:delta_n_M} and \eqref{eq:dwell_n_M} we can write:
\begin{equation}
    \Delta_{n+1} = \frac{1+2k-k^2}{(1-k)^2} \Delta_n - \frac{2k}{(1-k)^2} T .
\end{equation}
The solution for $\Delta_n$ with initial condition $\Delta_0$ is:
\begin{equation}\label{eq:delta_n_sol_alight}
    \Delta_{n} = \frac{T}{2-k} - \frac{T -  \Delta_0 \left(2 - k \right)}{2-k} \left(\frac{1 + 2 k - k^2}{(1-k)^2}\right)^n .
\end{equation}
The result is similar to the case where only boarding is allowed in Eq. \eqref{eq:delta_n_sol} except for the terms raised to the power $n$. Since $1 + 2 k - k^2 > 1$ for any valid $0<k<1$, introducing alighting decreases the distance between buses at a faster rate, causing bunching to occur sooner.
From this result, we can derive the expression for the number of loops before bunching $n^*$ by solving for $\Delta_{n^*} = 0$.
\begin{equation}\label{eq:n_1_alight}
    n^* = \frac{\log{\left(1 - \frac{\Delta_0}{T} \left(2-k\right)\right)}}
    {\log{\left( \frac{\left(1-k\right)^2}{1 + 2k- k^2}\right)}}.
\end{equation}
While Eqs. \eqref{eq:n_1} and \eqref{eq:n_M} are exact, Eq. \eqref{eq:n_1_alight} is based on the approximation $\bar{T}_n \approx T$ as discussed above.
A more precise solution to find the number of loops needed for bunching in the case where alighting is allowed can be found by explicitly iterating Eqs. \eqref{eq:delta_n_1} and \eqref{eq:dwell_n_1} considering $\bar{T}_n = T + \bar{\tau}_{n-1}^{(2)}$ until $\Delta_n \le 0$. The value $n^*$ counts the number of iterations. This approach requires an assumption for the value of $\bar{\tau}_{-1}^{(2)}$. The simplest option is to set $\bar{\tau}_{-1}^{(2)} = 0$, hence $\bar{T}_0 = T$, which ignores the the time spent by bus 2 at the destination bus stop in the first iteration. For the subsequent iterations however $\bar{T}_n = T + \bar{\tau}_{n-1}^{(2)}$ so the approximation is only at the very first step, the subsequent ones are exact. A better initial condition is $\bar{\tau}_{-1}^{(2)} = k/(1-k) \Delta_0$, where $\Delta'_{-1}$ is approximated as $\Delta_0$. We do not see a noticeable difference compared to the previous method for realistic values of $k < 0.1$, with differences within $0.1\%$.
Table \ref{tab1} compares the approximate Eq. \eqref{eq:n_1_alight} for $n^*$ to the number of loops before bunching obtained by numerically iterating Eqs. \eqref{eq:delta_n_1} and \eqref{eq:dwell_n_1} with initial condition $\bar{\tau}_{-1}^{(2)} = k/(1-k) \Delta_0$. As expected, Eq. \eqref{eq:n_1_alight} is an upper bound for the number of loops before bunching and the approximation is better for smaller initial distance $\Delta_0$ and arrival rate $s = k \times l$.

\begin{table}[htbp]
\caption{Bus bunching with alighting: comparison between the number of loops before bunching $n^*$ from Eq. \eqref{eq:delta_n_sol_alight} (first value) and numerical result (second value) for different $\Delta_0$ and $k=s/l$. The difference is also highlighted.}
\begin{center}
\begin{tabular}{|c|c|c|c|c|}
\hline

\textbf{} & \textbf{$k$=0.003}&  \textbf{$k$=0.009} & \textbf{$k$=0.027} \\
\hline
\textbf{$\Delta_0$ = 0.40T} & 134, 134 (0 \%) & 45, 45 (0 \%) & 15, 15 (0 \%) \\
\hline
\textbf{$\Delta_0$ = 0.45T} & 192, 191 (0.5 \%) & 64, 63 (1.6 \%) & 21, 20 (5.0 \%) \\
\hline
\textbf{$\Delta_0$ = 0.50T} & 543, 486 (12 \%) & 151, 132 (14 \%) & 41, 35 (17 \%)\\
\hline
\end{tabular}
\label{tab1}
\end{center}
\end{table}


\section{Waiting time for passengers}\label{sec:waiting}
Modelling the distance between buses is not only useful for predicting when buses will bunch. Under the assumption of constant arrival rate $s = k \times l$ of passengers at bus stops, $\Delta_n$ has information about the average waiting time for passengers at the bus stops. Firstly we consider the simplest case of a single origin bus stop (only boarding) as in section \ref{ssec:onestop}. Following the same idea as in \cite{vismara21}, the average waiting time for passengers at a bus stop arriving at a constant rate is half the waiting time of the passenger who waited for the longest. Assuming that passengers board ranked according to their arrival time at the bus stop (FIFO), the passenger who waited for the longest time at the $n$th loop is the first to arrive after the previous bus left the bus stop at the $(n-1)$th loop and consequently the first to board the new bus. Among passengers boarding the first bus, such waiting time is $\bar T_n - \Delta_n$, with $\bar T_n = T$. For passengers boarding the second bus, the longest waiting time is $\Delta'_n$,  hence the average waiting time for passengers of bus 1 and bus 2 is, respectively,
\begin{equation}\label{eq:wt12_1}
\begin{aligned}
\textit{WT}^{(1)}_n &= \frac{T - \Delta_n}{2} \\
\textit{WT}^{(2)}_n &= \frac{\Delta'_n}{2} = \frac{\Delta_n - k T}{2 (1-k)} .
\end{aligned}
\end{equation}
To calculate the average waiting time for all passengers during loop $n$, the quantities in Eq. \eqref{eq:wt12_1} have to be averaged with weight proportional to the number of passengers boarded. For bus $i$, the number of passengers boarded is proportional to the average waiting time $\textit{PB}^{(i)}_n = 2 s/(1-k) \times \textit{WT}^{(i)}_n$. The average waiting time for all the passengers boarded at the $n$th loop is:
\begin{equation}\label{eq:waiting}
\begin{aligned}
    &\textit{WT}_n = \frac{\textit{PB}^{(1)}_n \textit{WT}^{(1)}_n + \textit{PB}^{(2)}_n \textit{WT}^{(2)}_n}{\textit{PB}^{(1)}_n + \textit{PB}^{(2)}_n} = \\
    &\frac{\left( 2 - 2k + k^2 \right) \Delta^2_n + \left( 1-2k+2k^2\right) T^2 - \left(2 - 2k + 2k^2 \right) T \Delta_n}{\left( 1 - k\right) \left(k \Delta_n + \left( 1 - 2k\right) T\right)} .
\end{aligned}
\end{equation}
The value of $\Delta_n \in \left[ 0 ; 0.5\right]$ that minimises Eq. \eqref{eq:waiting} is $\Delta_n = 0.5 T$, i.e. perfectly staggered buses, for any value of $k \in (0, 0.5)$. For $k \ge 0.5$ the two buses would bunch immediately and the assumptions we made to reach this result are violated. In \cite{vismara21} and \cite{saw2020chaos} we show an expression for the average waiting time in the case of bunched buses.
In the limit of small $k$, the second line of Eq. \ref{eq:wt12_1} become $\textit{WT}^{(2)}_n \approx \frac{\Delta_n}{2}$ so $\textit{WT}^{(i)}_n$ are proportional to the headways between buses. Plugging this result in Eq. \ref{eq:waiting} shows that, in this regime of small $k$, $\textit{WT}_n \propto (T - \Delta_n)^2 + \Delta_n^2$, as found in \cite{Osuna1972} and \cite{Azfar2018}. The waiting time is, as before, minimised for $\Delta_n = 0.5 T$.
\section{Conclusion}
In this paper, we have presented a new analytical method to study bunching on a bus loop. For the three cases discussed in sections \ref{ssec:onestop}, \ref{ssec:Mstops} and \ref{ssec:alight}, we calculate when bunching occurs based on the initial distance between buses and the crowdedness of the bus stops. We show that increasing the number of bus stops while keeping the total demand constant delays bunching. In section \ref{sec:waiting}, we link the average waiting time with the distance between buses.
It is possible to extend the result to arbitrary scenarios of $M_o$ origin and $M_d$ destination bus stops but the set of equations for $\Delta_n$ require more initial conditions than $\Delta_0$, hence, as in the case of one origin and one destination bus stop in section \ref{ssec:alight}, a closed-form expression for $n^*$ is only approximate and the exact result is obtainable only by numerically iterating the equations from the initial conditions. Furthermore, with more destination bus stops, the formulae for $\Delta_n$ must account for the destination of passengers while computing alight times $\bar{\tau}_n$.
The calculations presented are for idealised systems, further analysis with realistic simulations are needed to validate and expand the results to real-world scenarios.

\bibliographystyle{unsrt}
\bibliography{mybib}
\end{document}